\documentclass[submission,copyright,creativecommons]{eptcs}
\providecommand{\event}{ICLP 2025 Technical Communications} 

\usepackage{iftex}
\usepackage{amsmath}
\usepackage{graphicx}
\usepackage{multirow}
\usepackage{url}
\usepackage{booktabs}
\usepackage{float}
\ifpdf
  \usepackage{underscore}         
  \usepackage[T1]{fontenc}        
\else
  \usepackage{breakurl}           
\fi

\newcommand{\constantSet}{\mathcal{C}}

\newcommand{\predicateSet}{\mathcal{P}}
\newcommand{\variableSet}{\mathcal{V}}

\def\domainSet{\mathcal{D}}

\def\assignAsset{\texttt{assign}_\mathbf{a}}
\def\assignFacil{\texttt{assign}_\mathbf{f}}
\def\useConstraint{\texttt{use}_\mathbf{c}}
\def\useOpt{\texttt{use}_\mathbf{o}}
\def\score{\texttt{score}}

\def\evac{\texttt{evac}}

\def\insults{\texttt{insultsAvailable}}
\def\vitals{\texttt{vitalsAvailable}}

\def\use{\texttt{use}}

\title{Reasoning about Medical Triage Optimization with\\Logic Programming}
\author{Jaikrishna Manojkumar Patil
\institute{Syracuse University\\ Syracuse, NY, USA}
\email{jpatil01@syr.edu}
\and
Adam Chapman
\institute{SimWerx\\
Denver, CO, USA}
\email{adam@simwerx.com}
\and
Richard Knuszka
\institute{SimWerx\\
Denver, CO, USA}
\email{rich@simwerx.com}
\and
John Chapman
\institute{SimWerx\\
Denver, CO, USA}
\email{john@simwerx.com}
\and
Paulo Shakarian
\institute{Syracuse University\\ Syracuse, NY, USA}
\email{pashakar@syr.edu}
}

\def\titlerunning{Reasoning about Medical Triage Optimization with Logic Programming}
\def\authorrunning{J.M. Patil, A. Chapman, R. Knuszka, J. Chapman \& P. Shakarian}
\def\copyrightholders{Patil, Chapman, Knuszka, Chapman \& Shakarian}
\begin{document}
\maketitle

\begin{abstract}
We present a logic programming framework that orchestrates multiple variants of an optimization problem and reasons about their results to support high-stakes medical decision-making. The logic programming layer coordinates the construction and evaluation of multiple optimization formulations, translating solutions into logical facts that support further symbolic reasoning and ensure efficient resource allocation—specifically targeting the “right patient, right platform, right escort, right time, right destination” principle. This capability is integrated into GuardianTwin, a decision support system for Forward Medical Evacuation (MEDEVAC), where rapid and explainable resource allocation is critical.  
Through a series of experiments, our framework demonstrates an average reduction in casualties by $35.75\%$ compared to standard baselines. Additionally, we explore how users engage with the system via an intuitive interface that delivers explainable insights, ultimately enhancing decision-making in critical situations. This work demonstrates how logic programming can serve as a foundation for modular, interpretable, and operationally effective optimization in mission-critical domains.
\end{abstract}


\section{Introduction}
Traditional medical triage algorithms, such as the Glasgow Coma Scale (GCS), the New Injury Severity Score (NISS), the Revised Trauma Score (RTS), and the LIFE score, efficiently categorize and prioritize patient care based on factors like injury severity, mental status, and vital signs ~\cite{teasdale1974assessment, osler1997modification, hr1989revision, sigle2023development}.  However, while these medical triage algorithms help efficiently categorize and prioritize patient care near the point of injury or at casualty collection points, these algorithms do not account for the strategic and logistical challenges critical to Forward Medical Evacuation (MEDEVAC)~\cite{bricknell} operations, particularly in high-intensity military conflicts like the 2022 Russo-Ukraine war.  Current MEDEVAC mission planning remains largely manual, making it difficult to scale in complex operational environments. Effective mission planning must consider factors such as geographic location, asset availability, proximity to care facilities, environmental conditions, and real-time operational intelligence. Given these challenges, the principle of ``right patient, right platform, right escort, right time, right destination'' becomes paramount~\cite{bricknell}. For example, a single patient triaged as ``Immediate'' might consume resources that could be better distributed to save multiple personnel categorized as “Urgent.” This underscores the need for a system that optimizes resource allocation and adjusts dynamically to operational changes. While optimization techniques are well-suited for addressing these challenges, there are questions about which type of optimization criteria to select~\cite{april2022descriptive,johnson2022development}, how to formulate the problem for a given scenario, and how to decide on a proper allocation of resources given a choice of solvable optimization problems. These gaps motivate the use of logic programming, which enables symbolic reasoning over constraints, preferences, and multiple optimization formulations in a structured and explainable manner.



In this paper, we introduce a logic programming based framework that allows for the \textit{orchestration} of one or more optimization problems for medical triage.  By creating a logical language to not only model medical triage situations but also to frame and reason about the solutions to optimization problems, we can create various optimization problems on the fly with the logic, send the results to an external solver, and then convert the results back into logical facts that allows for a form of ``what-if'' reasoning about the outcome if resource decisions are made in a certain manner.  In our paper, we describe how such reasoning can be done either via displaying results back to the user, or in automated fashion by reasoning over optimization problem outcome(s).  
To operationalize this framework in real-world mission planning, we extend our GuardianTwin triage platform, which integrates real-time data and decision-making algorithms to support Forward MEDEVAC coordination. GuardianTwin converts optimization outputs into intuitive interface elements, helping field medics and mission planners make informed decisions under pressure.
We then provide a suite of results where we examine the performance of the approach for cases where single and multiple optimization problems are formulated and show this provides on average at least $35.75\%$ reduction in casualties over standard baselines.  We also show how our approach can reason over multiple optimization problems as well as how it can be employed in time-evolving situations. 
While our primary focus in this paper is on Forward MEDEVAC operations, the logic programming framework we present is generalizable to other domains that require dynamic resource prioritization under constraints. The core idea of using logic rules to orchestrate optimization problem generation, reasoning over solutions, and adapting plans over time applies to any setting where explainability and constraint handling are essential. As such, our framework offers a reusable and interpretable foundation for decision support in time-critical domains beyond the military context.

\section{Background}\label{sec:background}
In this section we review patient criticality scoring (a feature included in the current version of GuardianTwin) and describe current thinking about medical triage with resource constraints.
\vspace{3pt}

\noindent\textbf{Patient Criticality Scoring.}  Current triage algorithms, such as NISS, RTS, and LIFE, have evolved to provide quantitative assessments of patient severity based on vital signs, injury scales, and other critical factors. NISS ranges from 0 (minor injuries) to 75 (severe injuries) and is based on the Abbreviated Injury Scale(AIS) for body regions, considering multiple injuries in a casualty. RTS ranges from 0 (most severe) to 12 (least severe) and incorporates vital signs such as systolic blood pressure, respiratory rate, and GCS. The LIFE score, ranging from 0 (most severe) to 100 (least severe), combines both AIS and vital information to provide a comprehensive severity score. In GuardianTwin and in this work, we have developed a simplified LIFE scoring system that equally weighs NISS and RTS scores to enhance decision-making in complex scenarios. Furthermore, to integrate multiple ranking systems, we normalize each score within a range of 0 (least priority) to 1 (highest priority). This normalization allows for consistent comparison across different scoring systems, accommodating situations where complete information might not be available for every casualty.  In this paper, we will use the notation $life$, $rts$, $niss$ to denote functions that accept a vector representation of a patient and return a normalized score as described above.
\vspace{3pt}
\begin{figure}[ht]
\centering
\includegraphics[width=0.49\columnwidth]{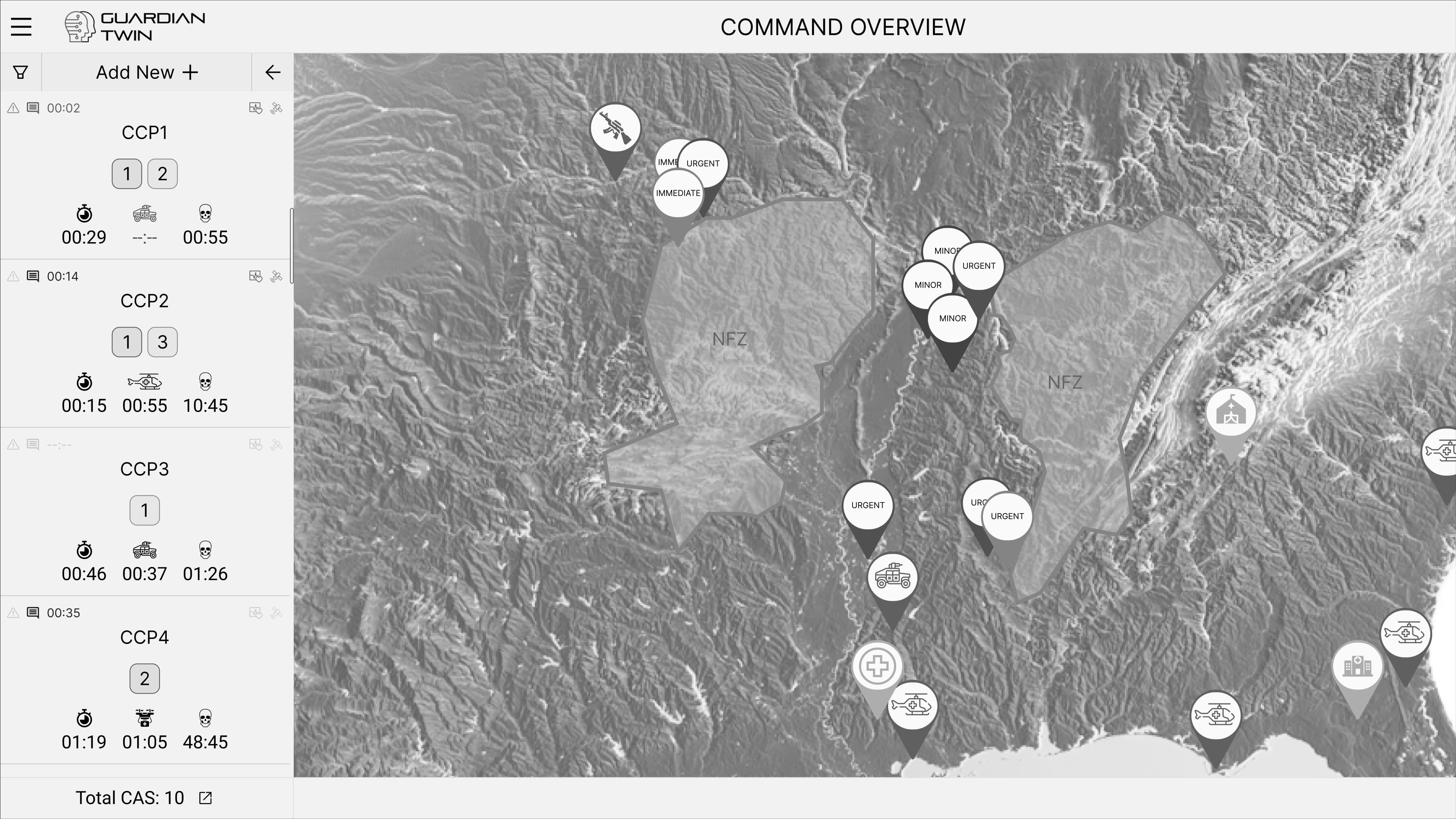} 
\hfill
\includegraphics[width=0.49\columnwidth]{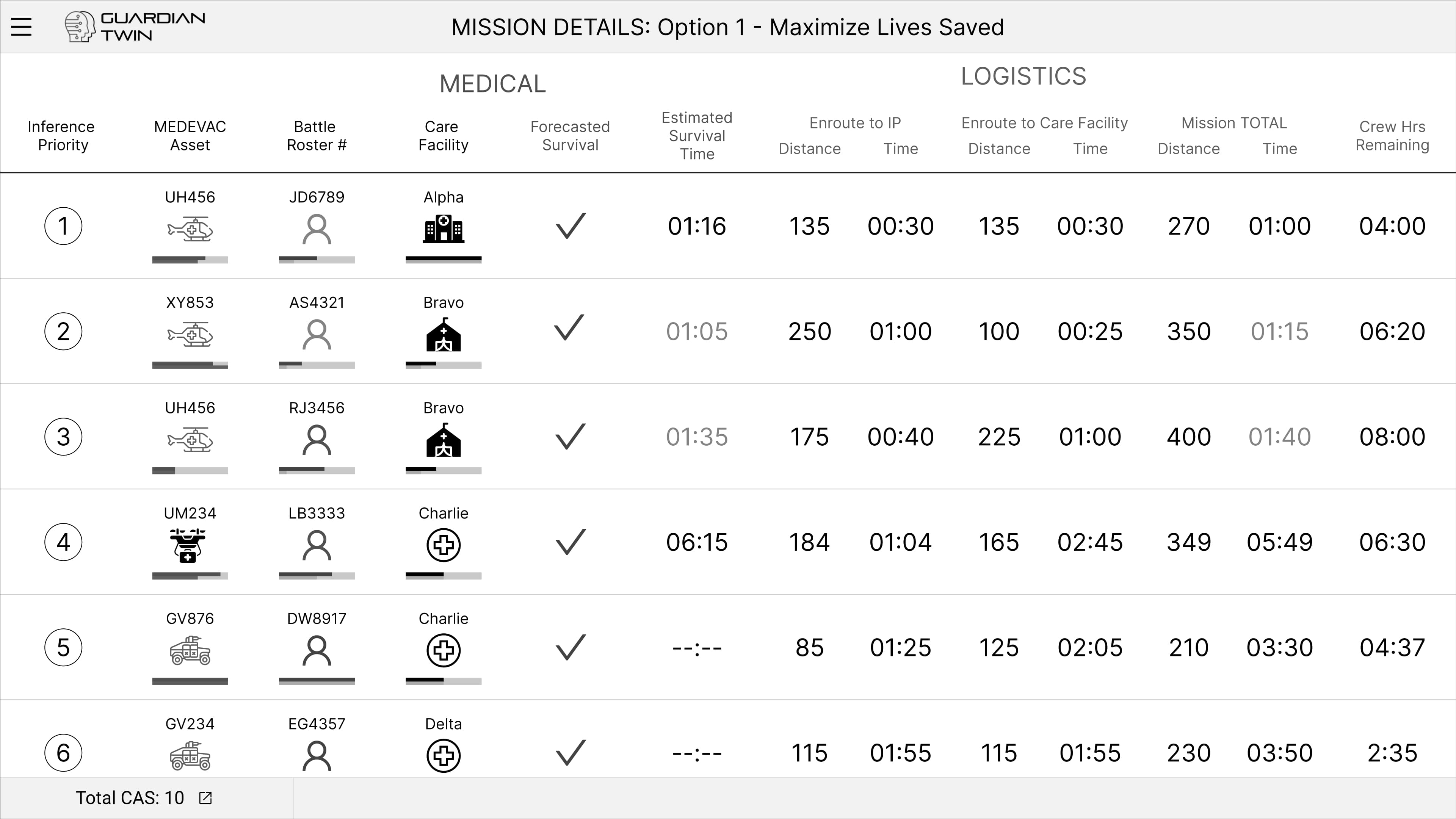} 
\caption{Left: GuardianTwin Command view.  Right: GuardianTwin Mission Details view (prototype).}
\label{fig:guardian-twin-ui}
\end{figure}

\noindent\textbf{Optimizing Triage with real world constraints.}  To illustrate the challenges in MEDEVAC mission planning, we present a scenario (see Figure~\ref{fig:guardian-twin-ui}(left)) with 5 assets and 11 casualties dispersed over a large geographical area: 3 ``Immediate,'' 5 ``Urgent,'' and 3 ``Minor.'' If asset allocation follows these categories, 3 ``Immediate'' and 2 ``Urgent'' casualties would receive resources, but this approach overlooks factors like response time, airspace control, and logistical considerations, potentially resulting in additional casualties. 
Given these complexities, framing mass casualty triage as an optimization problem becomes essential. However, the criteria for optimization depend on the specific situation. \textit{Urgency-based triage} prioritizes immediate intervention. Meanwhile \textit{reverse triage} focuses on return-to-duty time and \textit{situational triage} integrates operational constraints~\cite{april2022descriptive,johnson2022development}. 

Furthermore in MEDEVAC operations, ~\cite{jenkins2020robust} optimized Mobile Aeromedical Staging Facilities (MASFs) deployment and helicopter allocation based on logistical factors.  They also focus on logistical considerations like coverage of MASF and capacity of helicopters. Similarly, OPTEVAC model~\cite{sundstorm96} uses linear programming model to determine the optimal quantity and placement of evacuation assets based on the factors like available assets, area of battlefield, and preferred locations of medical care facilities. However, these models lack the framework to orchestrate multiple optimization problems and provide explainable traces that would help for taking next steps. Our logic-based framework bridges this gap, enabling decision support by integrating factors like distance to care facilities, airspace control, and MEDEVAC asset availability.
\vspace{3pt}

\noindent\textbf{Overview of GuardianTwin System.} GuardianTwin is a decision support platform designed mainly to optimize the survivability and resource management of battlefield casualties, particularly in large-scale combat operations (LSCO). GuardianTwin integrates real-time data, predictive modeling, and logic programming to maximize casualty survival and streamline resource management. The system creates virtual models of real-world entities such as casualties, assets, and care facilities. Due to the sensitive nature of this domain, it is not possible to get real-world data. However, GuardianTwin generates exercise-grade datasets specifically engineered for military exercises. These datasets are designed to reflect realistic casualty models, operational constraints, and logistical challenges, ensuring the relevance of platform to real-world applications. This data has been developed independently by GuardianTwin framework and is not a contribution of this work.
GuardianTwin serves three primary end-user groups: field medics at the Casualty Collection Point (CCP), Personnel Recovery Coordination Cell (PRCC) duty officers, and Patient Evacuation Coordination Cell (PECC) duty officers. It offers real-time triage support, predictive analytics, and resource prioritization to enhance decision-making in critical conditions. PRCC and PECC officers use the system to manage personnel recovery and evacuation planning, respectively. 
\vspace{3pt}

\noindent\textbf{Related Work.} Various approaches use reinforcement learning (RL) to optimize medical triage operations~\cite{yang2020,healthcare8020077} including in military medical triage~\cite{gelbard2024enhancing,RODRIGUEZ2023119751}. Similar RL approach is also used in humanitarian logistics for resource allocation~\cite{YU2021114663}. However, these approaches consider optimization criteria established ahead of time, which does not enable the user to understand the trade-off of various criteria (nor establish their own criteria for such trade offs in a logic program).  Other approaches include various machine learning models ~\cite{raita2019emergency,coslovsky2015clinical,fernandes2020clinical} which like the reinforcement learning approach all consider a single optimization criteria.  Medical triage is also studied in the field of AI for issues concerning recognition of casualties for robotic systems~\cite{kotha2023artemis}, and fairness in the triage process \cite{ghanbari2021fair}.  These issues do not consider the automated formulation of the triage problem as we do, but at the same time represent important concerns that would be worth considering in extensions to the framework described in this paper.
\section{Triage Logic and UI Prototypes}
\label{sec:prelims}
\begin{figure}[t]
\centering
\includegraphics[width=0.98\columnwidth]{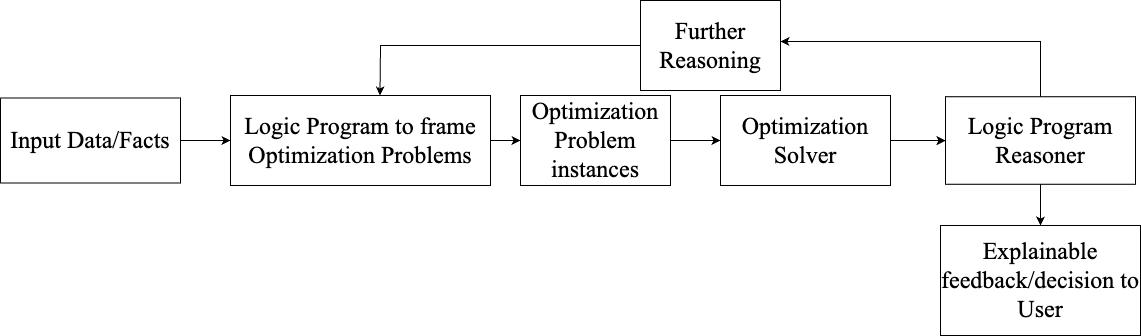} 
\caption{Overview of the framework for logic based orchestration of medical triage optimization.}
\label{fig:flow-overview}
\end{figure}

In this section we describe how we use logic programming to orchestrate one or more medical triage optimization problems as well as show examples of how users interact with the logic program via the GuardianTwin user interface.   We show an overall diagram of this approach in Figure~\ref{fig:flow-overview}. The key idea is that initial facts (including facts concerning the setup of the logic program) are input into the system, resulting in a deductive process that produces one or more optimization problems that are solved by an external solver.  The results of the solver are then re-interpreted as facts and added back into the logic program, allowing for further automated reasoning and/or analysis by the user.

\vspace{3pt}
\noindent\textbf{Logical Formalism.}  In this section, we briefly describe our logical language used for describing casualty triage based on a first order logic using datalog-style rules~\cite{aditya2023pyreason,ssTAI22}.  We assume a set of constant, variable, and predicate symbols (resp., $\constantSet$, $\variableSet, \predicateSet$) where the set of constants is divided into domains (denoted with the symbol $\domainSet$ often with a subscript).  Three key domains of importance we shall denote $\domainSet_{\mathbf{p}},\domainSet_\mathbf{a},\domainSet_\mathbf{f}$ for the sets of constants representing the sets of constants associated with geolocated casualties (personnel), assets, and care facilities respectively and are understood in the context of the MEDEVAC problem set.  Two other sets of constants, $\domainSet_\mathbf{o},\domainSet_\mathbf{c}$ are associated with optimization criteria and constraints respectively.  These constants are used for allowing the reasoning system to formulate an optimization problem.


Additionally, we denote the set of predicates as $\predicateSet$ and such predicates with terms (constants or variables) form atoms in the usual way.  For sake of brevity, we will only mention a few specific ones here.  There are three resulting from our overall process that describe the output - and actually are used in facts created by optimization problems orchestrated by the reasoning process.  First the unary predicate $\evac$ denotes if a casualty has been evacuated (e.g. $\evac(p)$ means that $p$ was evacuated). Next, the binary predicates $\assignAsset, \assignFacil$ take a casualty as the first argument and an asset (resp. facility) as the second - so $\assignAsset(p,a)$ means that casualty $p$ was assigned asset $a$ and $\assignFacil(p,f)$ means casualty $p$ was assigned to facility $f$.  For the reasoning process that stages the optimization problem, there is the binary predicate $\score$ that accepts an element from $\domainSet_\mathbf{p}$ as the first argument and a number in $(0,1)$ as the second.  In our framework, we allow this predicate to only appear in rule heads, as it is used to determine the type of score assigned to a casualty based on available data.  There are two unary predicates that are also used to help stage the optimization problem.  They are $\useConstraint$, $\useOpt$ and take constants from $\domainSet_\mathbf{c},\domainSet_\mathbf{o}$ respectively.  Here, $\useConstraint(c)$ means that we must enforce constraint $c$ during an optimization procedure and $\useOpt(o)$ means that we must use optimization criteria $o$.  Note that for only one $o \in \domainSet_\mathbf{o}$ can $\useOpt(o)$ be true when using a single optimization problem.  If we are using the framework to explore multiple optimization problems (as described later in the experiments), we extend the aforementioned predicates to binary predicates (we omit the details here for brevity).

We note that predicates like $\useConstraint,\useOpt$ are a direct result of the initial deductive process (which results in the setup of one or more optimization problems), while atoms formed with predicates $\evac$, $\assignAsset$, $\assignFacil$ are actually created as facts after one or more optimization problems are solved (for the case where we have more than one problem, we extend these predicates with an additional argument, a natural number $i$ indicating the $i$th optimization problem result).  For the results of a single optimization problem, we have created a ``Mission Details'' view screen for GuardianTwin (see Figure~\ref{fig:guardian-twin-ui}(right))summarizing which atoms formed with $\evac$, $\assignAsset$, $\assignFacil$ have resulted from a given optimization problem.


\begin{figure}[ht]
\centering
\includegraphics[width=0.49\columnwidth]{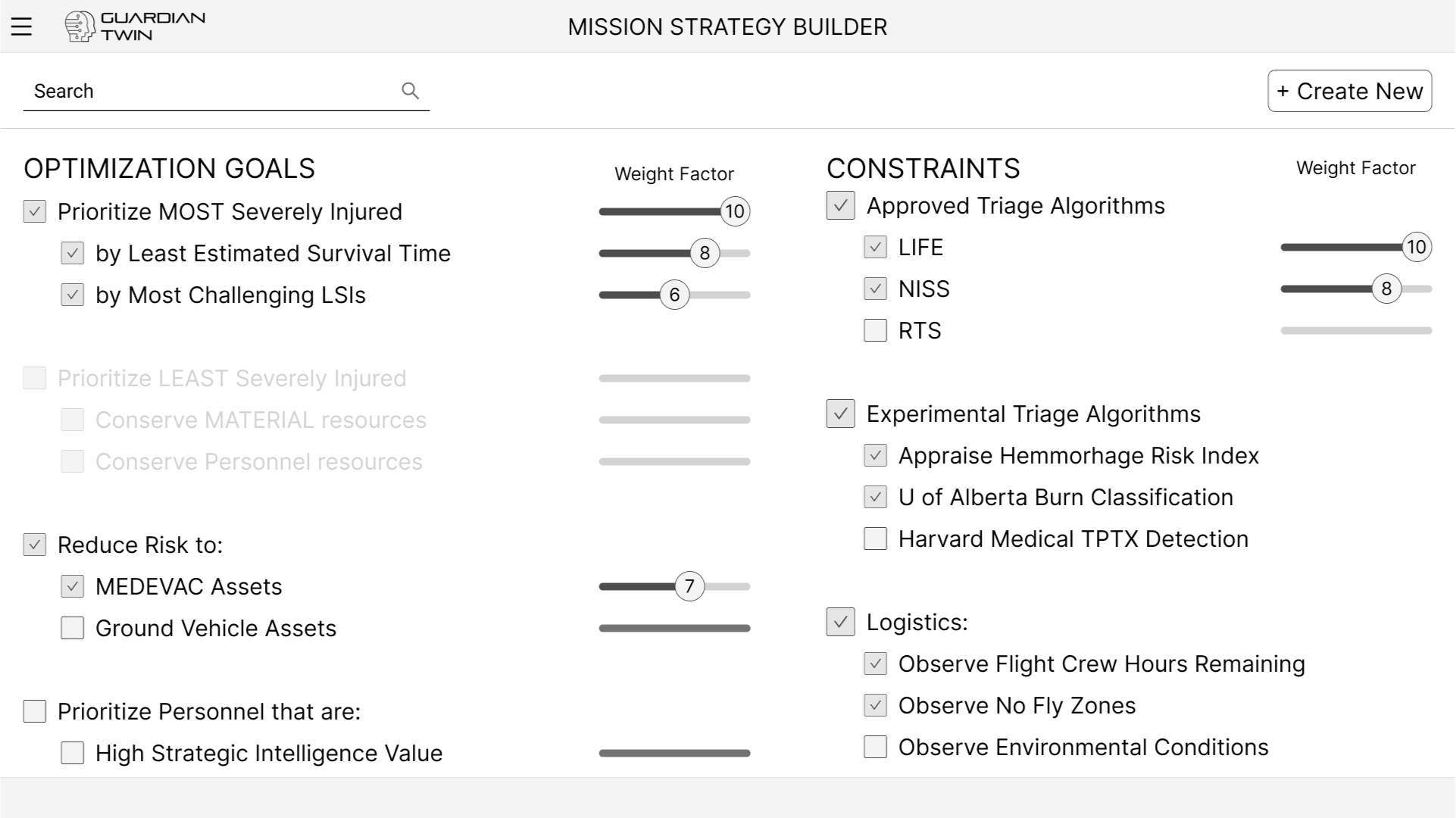}
\hfill
\includegraphics[width=0.49\columnwidth]{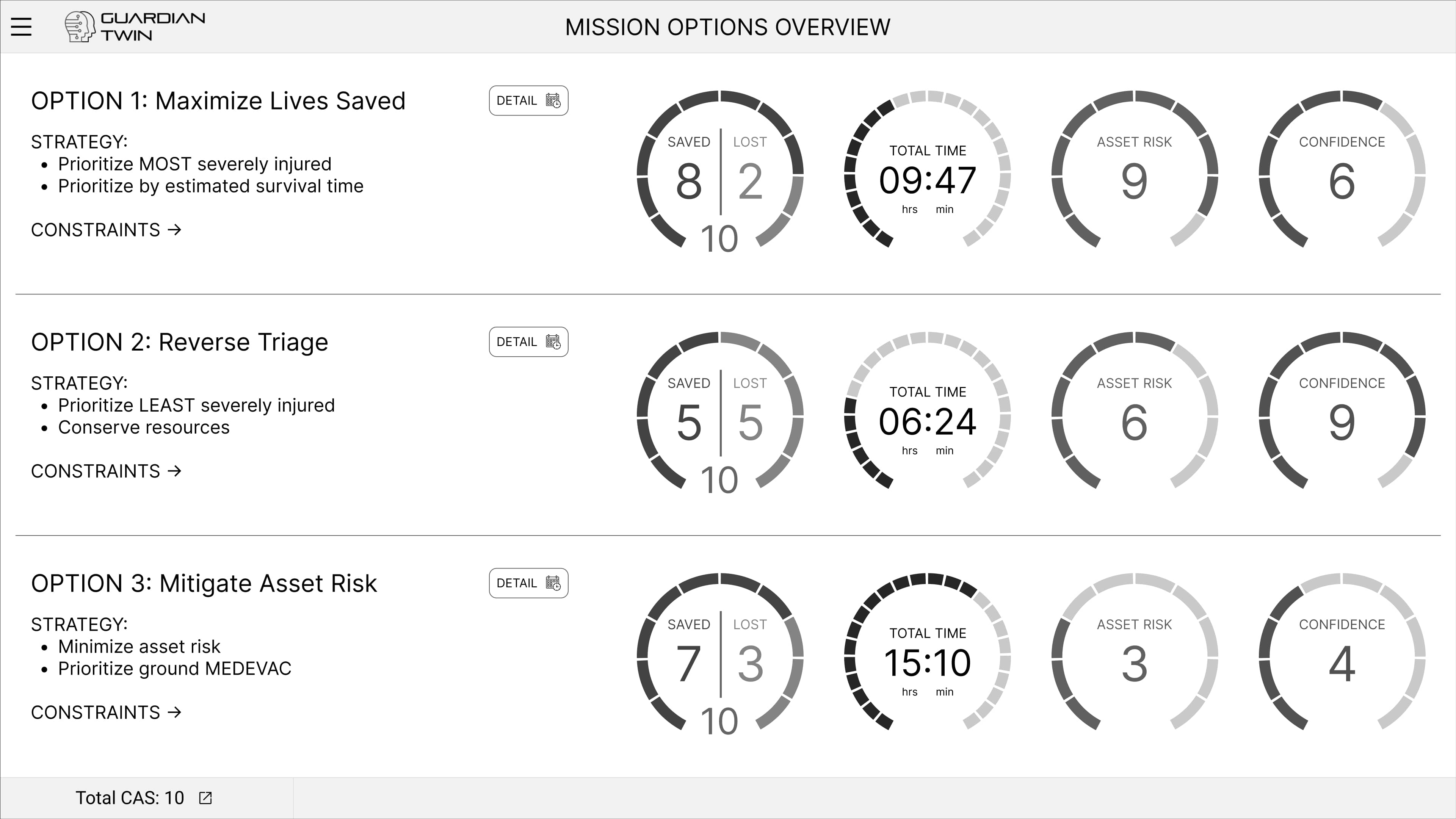} 
\caption{Left: GuardianTwin Strategy Builder view.  Right: Example results for three different optimization problems viewed in a prototype GuardianTwin interface.}
\label{fig:guardian-twin-ui-2}
\end{figure}

We will use strong negation ($\neg$) to denote a given atom is false (as opposed to merely ``not true'') - as our underlying logic allows uncertainty as a default truth value.  We will use the term ``facts'' to refer to a set of atoms or negations thought to be initially true.  We will also use atoms and negations in logic program rules (formal syntax is described in \cite{ssTAI22}).  A \textit{logic program,} often denoted as $\Pi$ in this paper is simply a set of facts and rules.  Unless mentioned explicitly, we shall assume that this logic program is fixed.  Table~\ref{tab:example-frame-opt} provides an example with the first rule states if \textit{insult severity scores} are \textit{available} for the casualty and \textit{vitals} is \textit{not available}, then we will use NISS scoring to compute triage score. On the other hand, second rule states if \textit{insult severity scores} are \textit{not available} for the casualty and \textit{vitals} is \textit{available}, then we will use RTS scoring to compute triage score. Third rule states that we will use LIFE scoring if \textit{both} insult severity score and vitals are \textit{available}.

\begin{table}[ht]
\centering
\caption{Logic program for framing optimization problems based on different criteria}
\label{tab:example-frame-opt}
\begin{tabular}{p{0.8\linewidth}} 
\toprule
\midrule
$\score(P,niss(P))\leftarrow \insults(P), \neg \vitals(P)$\\
\midrule
$\score(P,rts(P))\leftarrow \neg\insults(P), \vitals(P)$\\
\midrule
$\score(P,life(P))\leftarrow \insults(P), \vitals(P)$\\
\midrule
$\use(o_{rtd})\leftarrow$\\
\midrule
\bottomrule
\end{tabular}
\end{table}
For deductive inference with logic program $\Pi$, several different paradigms are possible. We refer the reader to references such as \cite{ks92,ssTAI22,aditya2023pyreason} for more complete review of syntax and semantics we use in this work. We selected this particular variant of logic programming as it allows for the representation of uncertainty.  Also, as per the aforementioned work, we assume initial values of all atoms or negations to be ``uncertain'' if not specified in an initial set of facts.  We assume that the logic is monotonic in that the truth value of ground atoms starts out as uncertain and moves to either true or false - but cannot change further (attempts to do so will result in an inconsistency as per \cite{ssTAI22}).  We will assume that the logic program in this work is consistent (and note in the above framework, the results of \cite{ssTAI22} tell us we can efficiently identify inconsistencies).  As a result our simple logic is amenable to a fixpoint-based deduction process (see the aforementioned references for details) that provides a set of all atoms and negations that can be derived from program $\Pi$ (either in the initial set of facts or derived from rules).  This process can be done efficiently as per \cite{ks92,ssTAI22} and implemented in PyReason~\cite{aditya2023pyreason} so we omit the details here.  For our purposes we care about the result of the deduction process.  As such is the case, for a given logic program $\Pi$, we will use the notation $\Gamma^*(\Pi)$ (a slight abuse of fixpoint notation used in \cite{ks92}) to refer to the set of all facts returned by the logic program.  Note that for the initial deductive process the key result of $\Gamma^*(\Pi)$ is to frame the optimization problems and perform subsequent reasoning.
\vspace{3pt}

\noindent\textbf{Using Logic Programming to Orchestrate Optimization Problems.}  Now we describe how we use this logical language to orchestrate optimization problems.  For starters, we define several functions over our set of constants.  In the background we already mentioned  $life$, $rts$, and $niss$ which map $\domainSet_\mathbf{p}$ to $(0,1)$.  Based on these, we define function  $scr: \domainSet_\mathbf{p} \to (0,1)$ defined as follows:
\begin{eqnarray*}
    scr(p) = \sup \{x \textit{ such that } \score(p,x) \in \Gamma^*(\Pi)\}
\end{eqnarray*}
We note that this function depends on $\score$ atoms deduced by $\Pi$ (recall that $\Pi$ is fixed and $\score$ only appears in rule heads).  The intuition here is that we now obtain a single scalar individual triage score for each casualty - and this can be derived from multiple rules (e.g., those in Table~\ref{tab:example-frame-opt}).  We note that unlike imputation, as used in traditional data mining, this approach allows for the logic around the triage score to reside in the logic program that is also used to formulate the optimization problem - which allows for more streamline integration with the user interface (e.g., the ``Strategy Builder'' depicted in Figure~\ref{fig:guardian-twin-ui-2} (left)) as well as allow better maintainability of the GuardianTwin codebase.  Other functions concerning return-to-duty time, and time required for life-saving intervention for a given casualty are denoted by $rtd : \domainSet_\mathbf{p} \to \Re^+$ and $lsi : \domainSet_\mathbf{p} \to \Re^+$.  These are defined in a similar fashion.  Additionally, the function $trip: \domainSet_\mathbf{p} \times \domainSet_\mathbf{a} \times \domainSet_\mathbf{f} \to \Re^+$, for some arguments $p,a,f$ (i.e., $trip(p,a,f)$ is the total trip time for casualty $p$ to facility $f$ via asset $a$.  Likewise functions $trip_\mathbf{p} : \domainSet_\mathbf{a} \times \domainSet_\mathbf{p} \to \Re^+$ and  $trip_\mathbf{f} : \domainSet_\mathbf{p} \times \domainSet_\mathbf{f} \to \Re^+$ represent travel times from the asset to casualty and casualty to care facility.  In orchestrating medical triage optimization, it is useful to specify constraints of different types (and, in fact these constraints are associated with elements of $\domainSet_\mathbf{c}$.  These constraints are applied on subsets of ground atoms \textit{after} the completion of the deduction process and we list several we use in this paper below.
\begin{eqnarray*}
    c_{lsi} := & \forall \assignAsset(p,a) \in \Gamma^*(\Pi): & trip_\mathbf{p}(a,p) \leq lsi(p)\\
    c_{scr}^k:= & \forall \evac(p) \notin \Gamma^*(\Pi): & scr(p) \leq k\\
    c_{rtd}^k:= & \forall \evac(p) \notin \Gamma^*(\Pi): & rtd(p) \geq k\\
    c_{air}^k := & \forall \assignAsset(p,a) \in \Gamma^*(\Pi): & trip_\mathbf{p}(a,p) \leq k \textit{ and}\\
    &\forall \assignFacil(p,f) \in \Gamma^*(\Pi): & trip_\mathbf{f}(p,f) \leq k
\end{eqnarray*}
Intuitively, constraint $c_{lsi}$ states that any casualty assigned an asset is provided that asset before life saving intervention is required.  Constraint $c_{scr}^k$ (defined for some constant $k$) ensures that no casualty not evacuated has a triage score greater than $k$ (and $c_{rtd}^k$ is simply the analog for return to duty).  Constraint $c_{air}^k$ (defined for some constant $k$) ensures that the time in the air for an asset on any leg of the journey does not exceed $k$ time units.  We note that not all of these constraints are necessarily to be enforced at once (and we may even have different constraints used for different optimization problems orchestrated by the logic program).  The key is that the logic program orchestrates the constraints properly (e.g., for a single optimization problem, the atom $\useConstraint(c_{lsi})$ included as a fact would indicate that the constraint $c_{lsi}$ should be used).

We can use the logic program to select the optimization criteria in a similar manner.  We define two optimization criteria that we use in this paper as follows.
\begin{eqnarray*}
    o_{scr} := & \sum_{\substack{p \textit{ such that}\\ \{\assignAsset(p,a), \\ \assignFacil(p,f)\} \subseteq \Gamma^*(\Pi)}}1+scr(p)-trip(p,a,f)\\
    o_{rtd} := & \sum_{\substack{p \textit{ such that}\\ \{\assignAsset(p,a), \\ \assignFacil(p,f)\} \subseteq \Gamma^*(\Pi)}}1+\frac{1}{1+rtd(p)}-trip(p,a,f)
\end{eqnarray*}
 Here, $o_{scr}$ is an objective function resembling urgency based triage in that we want to increase the number of personnel evacuated and prioritize a higher individual triage score, while attempting to reduce total travel time.  Likewise, $o_{rtd}$ is similar, except that it prioritizes return-to-duty (hence $scr(p)$ is replaced with $\frac{1}{1+rtd(p)}$.  Again, with a single optimization problem, $\useOpt(o_{rtd})$, as seen as a fact in the last line of Table~\ref{tab:example-frame-opt}, would mean that we use the reverse triage optimization criteria.  Note that again, with multiple optimization problems, we would use binary version of $\useOpt$ (with the second argument being a natural number for the associated optimization problem).  We note that the result of the deduction process gives us an optimization problem of the form ``optimize $o$ such that $\useOpt(o) \in \Gamma^*(\Pi)$ subject to all $c$ such that $\useConstraint(c) \in \Gamma^*(\Pi)$.''  We provide examples of how we use the logic to create integer programs in an online appendix~\footnote{https://tinyurl.com/appendixiclp2025}.
However, the procedure is straightforward, as all constraints and optimization criteria are linear combinations over the integers zero and one.

From a user perspective, setting the facts with predicates $\useOpt,\useConstraint$ will be entered in the GuardianTwin user interface in view called ``Startegy Builder'' shown in Figure~\ref{fig:guardian-twin-ui-2} (left)).  Based on the user's selections, facts formed with these two predicates are added for one or more optimization problems.
\vspace{3pt}


\noindent\textbf{Reasoning Over Solutions to Optimization Problems.}  The PyReason implementation for logic programming allows us to maintain the logic program in memory and even add additional facts to it.  As a result, we can follow the procedure of running the deductive process for  logic program $\Pi$ and for each $o$ such that there is some $\useOpt(o,i) \in \Gamma^*(\Pi)$ we solve the associated optimization problem, where the result can be translated into a set of facts (e.g., $\Pi_{o,i}$).  Then, we run the deduction process on that result that is equivalent to the following:
\begin{eqnarray*}
    \Gamma^*(\Pi \cup \bigcup_{\substack{o \textit{ such that there exists }i\ \textit{ where}\\ \useOpt(o,i) \in \Gamma^*(\Pi)}}\Pi_{o,i})
\end{eqnarray*}
We note that the new facts added (e.g., each set $\Pi_{o,i}$) will cause additional rules in $\Pi$ to fire to properly select among the solutions to the optimization problems.  For example, we may want to select an optimization problem that is within a certain percent of optimal of urgency based triage while still honoring as many constraints as possible.  Alternatively, the explainable trace of logic produced by the deductive process can also be shown to a user to allow them to make further decisions.  We have started prototyping user interface designed to communicate the results of different optimization problems back to the user in GuardianTwin (Figure~\ref{fig:guardian-twin-ui-2} (right))).
It is important to note that end users of GuardianTwin—such as field medics or coordination officers will not interact with logic programs or optimization parameters directly. Instead, they engage with the system through high-level interface components like the Strategy Builder Figure~\ref{fig:guardian-twin-ui-2} (left)), where abstract operational goals are translated automatically into logical facts. This will shield users from underlying complexity and ensure usability under high-stress conditions.


\section{Experiments}
\noindent\textbf{Experimental Setup.} Due to the sensitive nature of military operations, it is impossible to obtain real-world data for our experiments. To address this limitation, we evaluate our approach using an exercise-grade dataset created within the GuardianTwin framework. This dataset is engineered to capture realistic MEDEVAC operational scenarios by incorporating geographical constraints, casualty distributions, and asset limitations. The generated casualties, assets, and care facilities are designed to align closely with real-world parameters. The use of this exercise-grade dataset provides valuable insights into the potential performance of GuardianTwin in operational conditions. This dataset comprises three components: casualties, assets, and care facilities. Each casualty is represented as a tuple of the form $\langle \textit{name}, AIS_{1}, \ldots,  AIS_{n}, \textit{systolic blood pressure},$ $\textit{respiratory rate}, \textit{GCS}, \textit{location}\rangle$. Here $AIS_i$ is the severity score of insult \textit{i} with possible values in $\{0,1,2,3,4,5,6\}$. This vector representation of the constant allows us to easily implement many of the functions described in the previous section.  Likewise, assets are represented by tuples of the form $\langle \textit{name}, \textit{location}, \textit{range}, \textit{speed}, \textit{remaining crew duty hours} \rangle$ and care facilities are represented as tuples of the form $\langle \textit{name}, \textit{location}\rangle$. The locations of all casualties, assets, and care facilities are generated within a geographical area extending from the southern border of Arizona to the northern border of Utah, and from the western border of Arizona to the eastern border of Colorado. For each experimental run, we randomly select \( n \) casualties, \( k \) assets, and \( m \) care facilities from the provided dataset, with the goal of optimally assigning each casualty to an asset and care facility in order to maximize different criteria (as orchestrated by the logic program). To assess the advantages of our approach, we carried out multiple experiments to demonstrate abilities of our framework for MEDEVAC mission planning. We analyzed a sample of 250 data points. These data points were created using 10 samples from each of the following settings: \{\(n=25, m=10, k=1\)\}, \{\(n=25, m=10, k=2\)\}, \ldots \{\(n=25, m=10, k=25\)\}.  We utilized PyReason~\cite{aditya2023pyreason} to support logic programming inference and the mip package~\cite{mip} for integer program optimization.  Experiments were run on a MacBook equipped with 12-core CPU, 18-core GPU, and 18 GB unified memory. All optimization problems solved within an average of 6 milliseconds per scenario across varying numbers of assets and casualties, with none exceeding 10 milliseconds. The logic program deduction steps completed in 3 to 5 seconds on all runs, demonstrating practical tractability for near real-time decision support. The code and data used in the experiments is available at \url{https://github.com/jaikrishna-patil/guardian-twin-opt}.
\vspace{3pt}

\noindent\textbf{Exercise-Grade Data for Validation.} The sensitive nature of military operations makes it impossible to collect real-world data. Our evaluation overcomes this limitation by using exercise-grade datasets which were specifically created to simulate real-world MEDEVAC situations and thus offer a strong operational testing environment for our framework. The datasets combine realistic elements like casualty profiles with geographic constraints and asset capabilities. Casualties possess attributes that include injury severity and vital signs along with geographic location and urgency levels which are assessed through established scoring systems like NISS, RTS, and LIFE. The dataset contains assets with operational details such as range capabilities and speed specifications along with base locations and remaining duty hours while care facilities are placed geographically to match real-world locations. The exercise-grade environment evaluates spatial factors such as distance measurements, terrain obstacles, and time-sensitive limitations to match real-world operation requirements. Expert knowledge and research studies about military and humanitarian MEDEVAC activities shaped the dataset design to achieve realistic scenarios with different casualty volumes and operational challenges. This data is not a contribution to this work; it was developed separately to support military exercises. This framework achieves practical applicability through simulation of real-world complex constraints while maintaining theoretical robustness and paving the way for live deployment upon obtaining required permissions.
\vspace{3pt}

\noindent\textbf{Optimization Methods and Baselines.}  We use the logic program to orchestrate three different types of optimization problems which we investigate both individually and collectively by reasoning over different solutions.  For comparison purposes, we also describe three basline models.
\vspace{4pt}

\noindent\textit{Urgency based Optimization.}  This optimization problem uses optimization criteria $o_{scr}$, constraint $c_{lsi}$, and constraint $c_{scr}^k$ for values $k=\{0.5,0.6,0.7,0.8,0.9,1.0\}$.
\vspace{4pt}

\noindent\textit{Reverse Triage Optimization.}  This optimization problem uses optimization criteria $o_{rtd}$, constraint $c_{lsi}$, and constraint $c_{rtd}^k$ for values $k=\{50,40,30,20,10,0\}$.
\vspace{4pt}

\noindent\textit{Situational Triage Optimization.}  This optimization problem uses optimization criteria $o_{scr}$, constraint $c_{lsi}$, constraint $c_{air}^k$ for $k=1$
\vspace{4pt}

\noindent\textit{Random Baseline Model (B1).} In this baseline model, casualties are selected randomly and assigned to a randomly chosen available asset and care facility while adhering to constraint $c_{lsi}$.
\vspace{4pt}

\noindent\textit{Triage Priority Baseline Model (B2).} In this model, casualties are selected one by one in descending order of $scr$ and randomly assigned an available asset and care facility while adhering to constraint $c_{lsi}$.
\vspace{4pt}

\noindent\textit{Return to Duty Priority Baseline Model (B3).} This model prioritizes casualties by ascending order of $rtd$ and randomly assigned an available asset and care facility while adhering to constraint $c_{lsi}$.
\vspace{4pt}

These baselines were designed to mirror practical and interpretable heuristics inspired by real-world triage decision-making. While several reinforcement learning (RL) and machine learning (ML) approaches exist for triage and humanitarian logistics (see Section \ref{sec:background}), such methods are typically trained to optimize a fixed criterion, often without constraint reasoning or user-in-the-loop customization. As our logic framework supports constraint-based orchestration and multi-objective selection, direct empirical comparison to black-box learning methods is not straightforward. We therefore use structured internal baselines  that align with clinical heuristics.
\vspace{3pt}

\begin{figure}[ht]
\centering
\includegraphics[width=0.49\columnwidth]{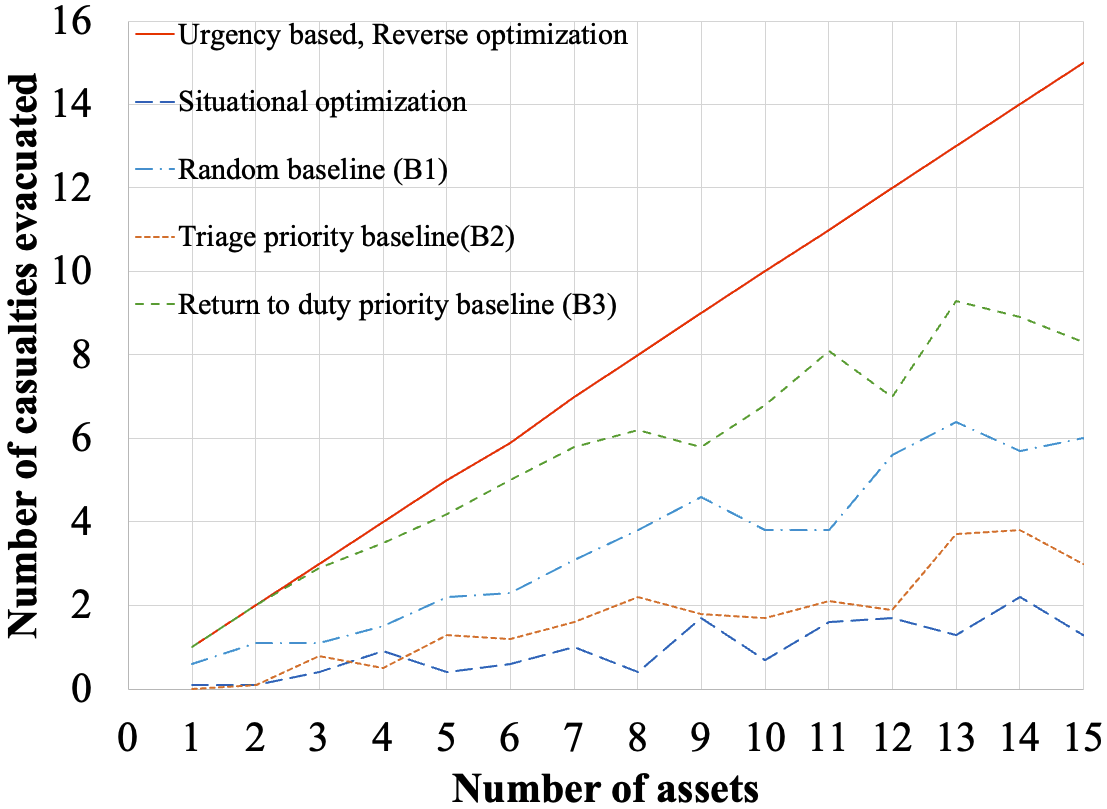} 
\includegraphics[width=0.49\columnwidth]{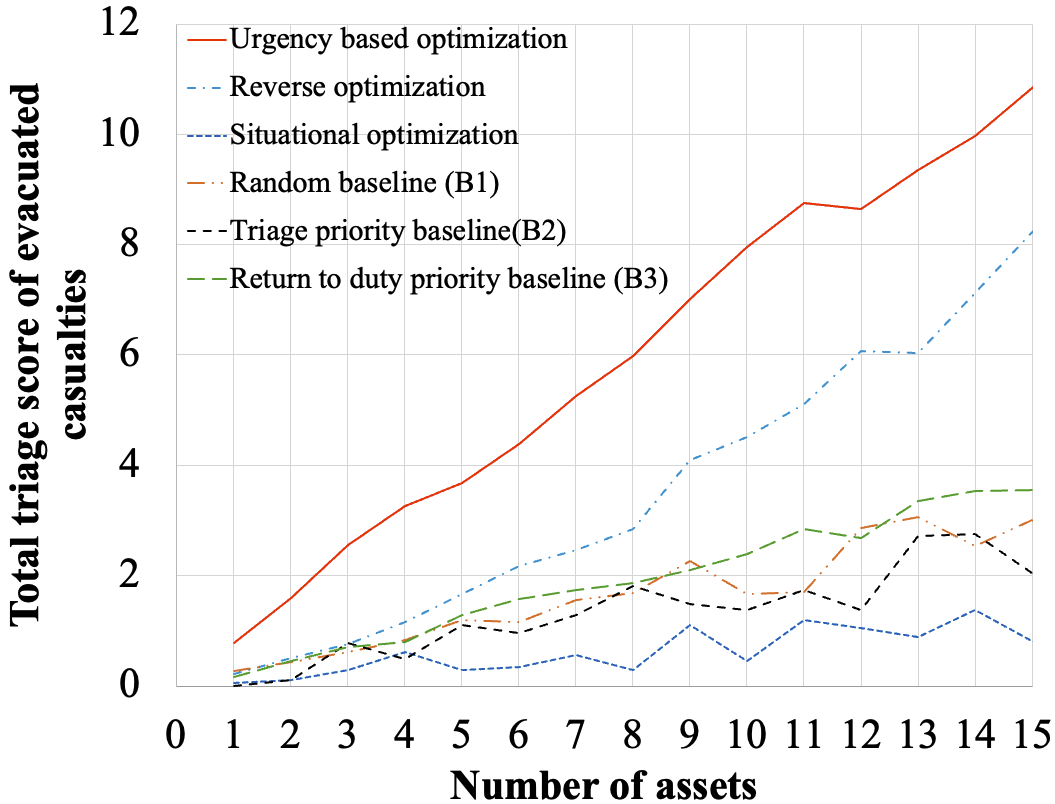} 
\caption{Number of casualties evacuated successfully (left) and total traige score of all evacuated casulaties (right) by single optimization problems and baseline models for different number of assets.}
\label{fig:baseline-num-cas-served}
\end{figure}

\noindent\textbf{Single Optimization Problem.}  Figure~\ref{fig:baseline-num-cas-served} (left) demonstrates that the assignments provided by urgency based and reverse optimization problems resulted in more casualties being evacuated than any of the baselines or situational optimization across nearly all asset levels (in that figure, we use $25$ casualties and $10$ facilities).  On average, urgency based and reverse optimization assignments evacuated more than twice the casualties evacuated by B1 and B2. This improvement goes to $61.32\%$ increase when compared with B3. We also observe that number of casualties evacuated by assignments provided by situational optimization problem underperformed relative to our other methods. This makes sense because latter optimization problem has additional air time constraint to mitigate asset risk. In Figure~\ref{fig:baseline-num-cas-served} (right) we examine the sum of the triage score for evacuated casualties.  Here, we also see the best performance is obtained with urgency based and reverse optimization - though they diverge more than in the earlier experiment.  

\begin{figure}[ht]
\centering
\includegraphics[width=0.49\columnwidth]{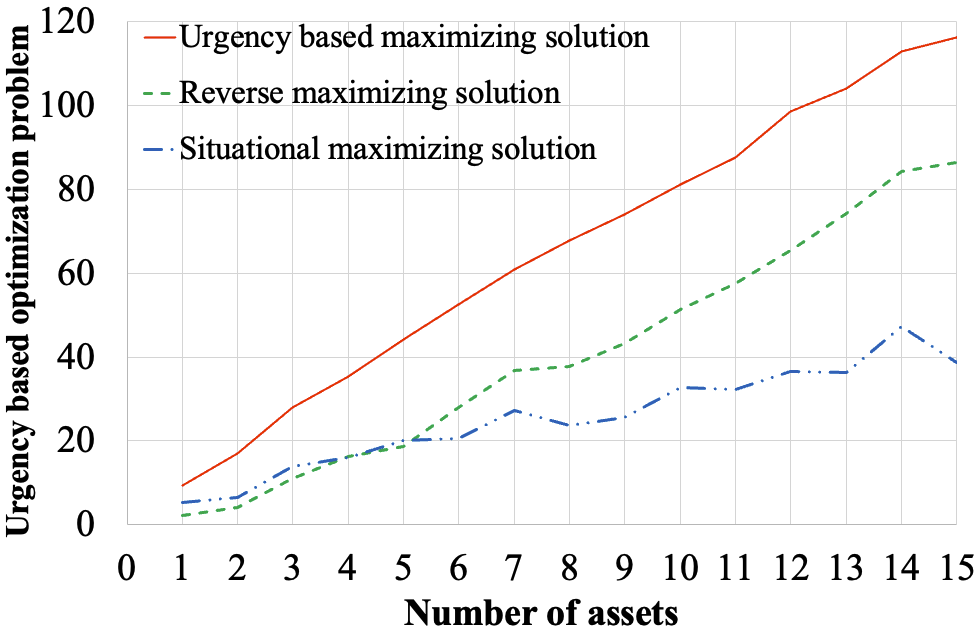} 
\includegraphics[width=0.49\columnwidth]{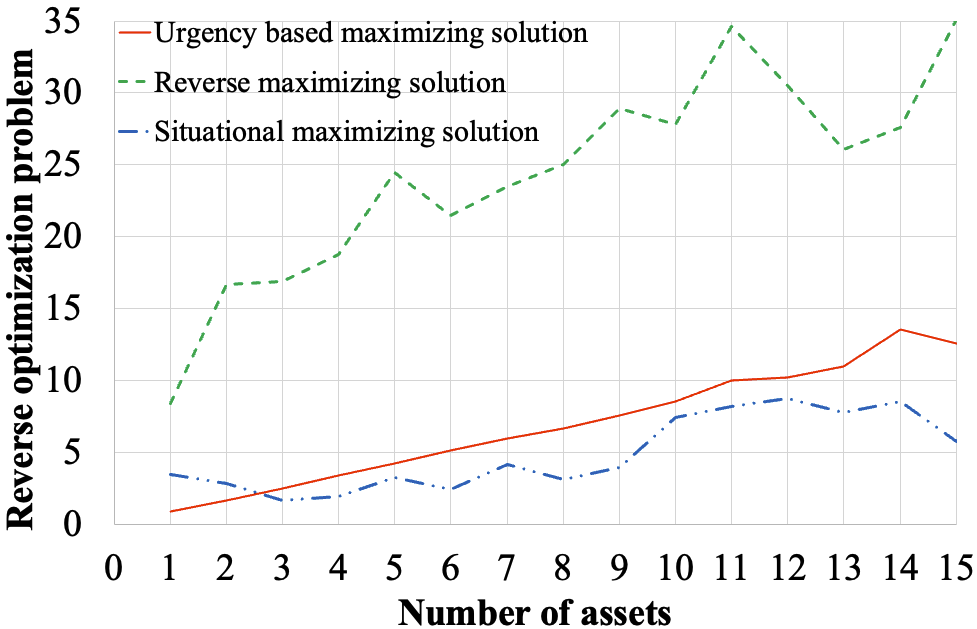} 
\caption{Performance of urgency based, reverse and situational optimization solution on urgency based optimization criteria (left) and reverse triage optimization criteria (right).}
\label{fig:multi-op}
\end{figure}

\begin{figure}[ht]
\centering
\includegraphics[width=0.49\columnwidth]{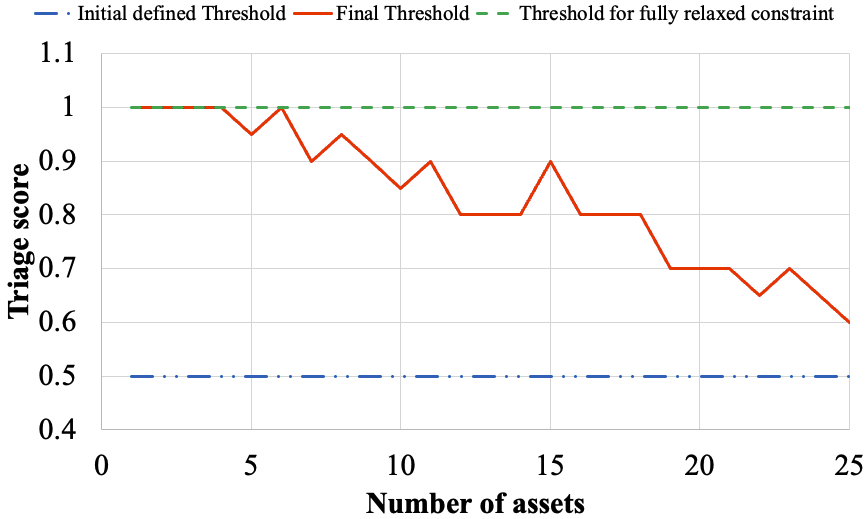} 
\includegraphics[width=0.49\columnwidth]{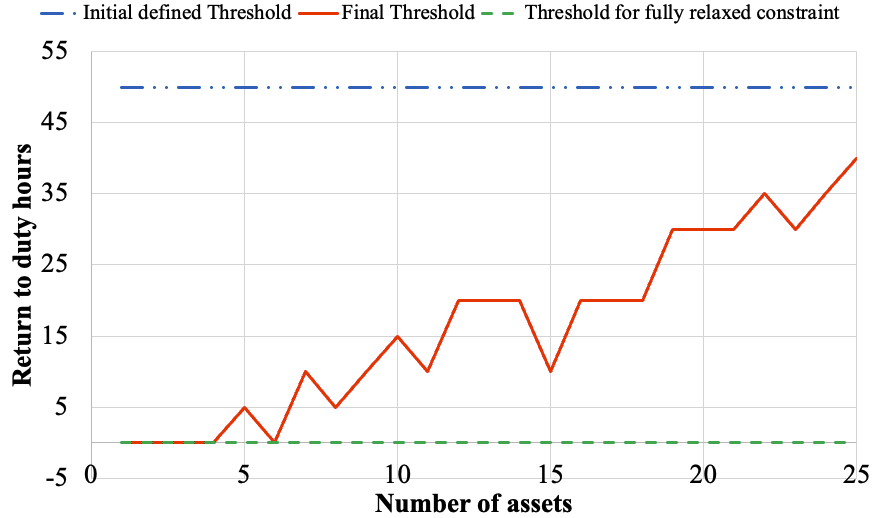} 
\caption{Left: Relaxation of constraint with triage score.  Right: Relaxation of constraint with return to duty hours.}
\label{fig:multiconstraint-score-threshold}
\end{figure}

\noindent\textbf{Multiple Optimization Problems.} A key aspect of our framework is the ability to examine the results of multiple optimization problems simultaneously.  We investigate this from two different respects.  First, we look at relative performance of an optimization problem with respect to other optimization criteria - which can provide insight into the development of a logic program to select among different optimization criteria.  The second deals with constraint relaxation or removal.  In this case, particularly with constraint that use a threshold (e.g. $c_{scr}^k$) we may not have a feasible solution to the optimization problem, so instead we relax constraints (via rules in logic program $\Pi$) and re-stage new optimization problems. In Figure~\ref{fig:multi-op} we examine the performance of a given optimization problem staged by $\Pi$ with respect to other optimization criteria.  In the left-hand panel, we see, for example, that the Reverse Triage Optimization problem comes within $66\%$ of the value of $o_{scr}$ for several asset ranges (e.g., for $7$ assets) while providing a 4-fold increase over the Urgency based Triage Optimization problem for $o_{rtd}$.  This type of understanding, can be, in-turn leveraged by the logic program after facts created from the optimization problem are placed back in the logic program.  A natural way to consider this is to have logical rules that relax or remove constraints.  In Figure~\ref{fig:multiconstraint-score-threshold} we consider this where we relax $k$ values for constraints $c_{scr}^k$ (left) and $c_{rtd}^k$ (right) using rules in $\Pi$ that reason about the results of multiple optimization problems.  We observe in both cases that for smaller number of available assets i.e. \(k=\{1,2,3,4,6\}\), constraint needs to be fully relaxed in order to get a feasible solution. However, as number of available assets increase, we are able to get feasible solutions while still adhering to a subset of the constraints.

\begin{figure}[H]
\centering
\includegraphics[width=0.9\columnwidth]{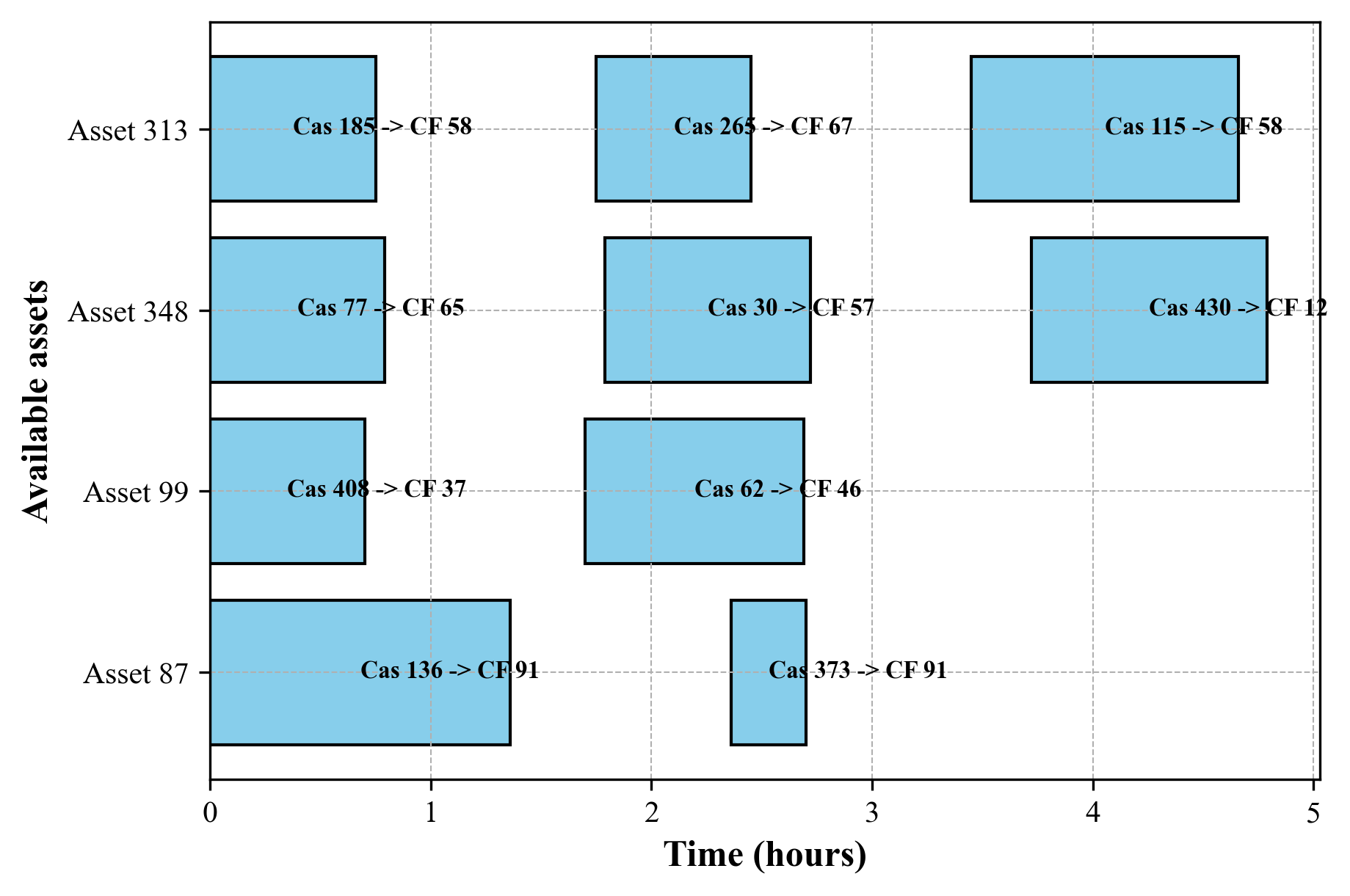} 
\caption{Sample Iterative schedule.}
\label{fig:gantt-chart-paper-style}
\end{figure}

\noindent\textbf{Orchestrating sequential optimization problems. } In real MEDEVAC scenarios, assets are assigned to casualties and then re-assigned upon completion of their mission.  As our framework allows for the iterative orchestration of optimization problems, we can leverage it to orchestrate sequences of optimization problems over time.  In Figure~\ref{fig:gantt-chart-paper-style} we show a Gantt chart resulting from this iterative process.  As part of future work, we look to leverage the temporal aspects of the underlying logic~\cite{aditya2023pyreason} to allow for additional modeling (e.g., changes to the crew over time, enemy situation, etc.) as well as explore the use of planning to examine optimal solutions over multiple time steps.

\section{Conclusion}
In this paper, we showed how we could use logic programming to orchestrate one or more optimization problems for medical triage resource allocation and then, in turn, reason about the results.  Such reasoning can be used to select the results of a particular optimization problem or to sequence multiple optimization problems over time.  We show how the symbolic representations of the logic program provide a natural analogue for a user interface - allowing the user to customize the logic program and stage optimization problems in various ways.  
As we continue the development of GuardianTwin and move toward deployment, we shall refine how the user can benefit from automated decisions to reason over the results of the optimization problems, explore trade-offs in medical triage, as well as incorporate additional knowledge such as data about the enemy and other characteristics of medical treatment.

\noindent\textbf{Path to Deployment.} Due to the sensitive nature of military MEDEVAC operations, obtaining real-world data or immediate access to live deployment scenarios is not feasible. Instead, GuardianTwin is evaluated using exercise-grade datasets generated within the framework itself, which are designed to simulate realistic casualty profiles, resource constraints, and geographic challenges. This approach enables rigorous, reproducible testing under operationally relevant conditions while adhering to security protocols. The deployment pathway follows a structured, three-phase strategy: initial validation using simulated scenarios and expert feedback; beta testing in controlled environments with field medics and coordinators; and eventual deployment in live operations, subject to required security clearances and institutional approval. 
This phased progression ensures both technical reliability and practical readiness for real-world use.

\bibliographystyle{eptcs}
\bibliography{iclp25}
\end{document}